\documentclass[a4paper,11pt]{article}
\pdfoutput=1 

\usepackage{jcappub} 

\usepackage[T1]{fontenc} 

\usepackage{url}
\usepackage{color}
\usepackage{comment}
\usepackage{mathrsfs}
\usepackage{hyperref}
\usepackage{braket}
\usepackage[dvipsnames]{xcolor}
\usepackage{natbib}
\usepackage{physics}
\usepackage{calc}
\usepackage{amsmath, amssymb}

\usepackage{here}
\usepackage{siunitx}
\usepackage{float}
\usepackage{url}
\usepackage{latexsym}
\usepackage{listings, jvlisting}
\usepackage{ascmac}
\usepackage{mathcomp}

\usepackage{minitoc}
\usepackage{amsmath}
\usepackage{amssymb}

\newcommand{\simgt}{\lower.5ex\hbox{$\; \buildrel > \over \sim \;$}}

\newcommand{\mr}{\mathrm}

\newcommand{\B}[1]{{\boldsymbol{#1}}}

\newcommand{\sqbra}[1]{
\left[
#1
\right]
}
\newcommand{\parbra}[1]{
\left(
#1
\right)
}

\title{Quadratic estimators for unwindowed power spectrum of galaxy-galaxy weak lensing and its application to $P_{\rm gm}(k)$ estimation}

\author[a,b]{Taisei Terawaki,}
\author[a,c]{Masahiro Takada,}
\author[a,b]{Takanori Taniguchi}

\affiliation[a]{Kavli Institute for the Physics and Mathematics of the Universe (WPI), The University of Tokyo Institutes for Advanced Study (UTIAS), The University of Tokyo, 5-1-5 Kashi- wanoha, Kashiwa-shi, Chiba, 277-8583, Japan}
\affiliation[b]{Department of Physics, The University of Tokyo, Bunkyo, Tokyo 113-0031, Japan}
\affiliation[c]{Center for Data-Driven Discovery (CD3), Kavli IPMU (WPI), UTIAS, The University of Tokyo, Kashiwa, Chiba 277-8583, Japan}

\emailAdd{taisei.terawaki@ipmu.jp}

\abstract{
Galaxy-galaxy weak lensing provides a powerful means of measuring the average matter distribution around lens galaxies -- i.e., the galaxy bias relation. Properly accounting for the spin-2 nature of weak lensing distortions, we develop a quadratic estimator for measuring the $E$- and $B$-mode angular power spectra from galaxy-galaxy weak lensing, correcting for survey window effects arising from, for example, survey geometry and bright star masks. The estimator can be implemented efficiently by adopting FFTs on pixelized maps of the lens galaxy distribution and source galaxy ellipticities, under the flat-sky approximation. Using simulated weak lensing fields and halo catalogs in the light-cone ray-tracing simulations, we show that the estimator can recover the underlying $E$-mode power spectrum, $C_{{\rm g}E}(\ell)$, to within a few percent in fractional error, while minimizing the leakage of $E$-mode into the $B$-mode power spectrum, in each multipole bin over the wide range of multipoles (up to $\ell \sim 3000$ studied in this paper). We then discuss that the estimator can be used to estimate the 3D galaxy-matter power spectrum, $P_{\rm gm}(k)$, by dividing lens galaxies into multiple redshift slices. We also derive an optimal weighting for each lens redshift slice in the shot noise-limited regime for the estimation of $P_{\rm gm}(k)$, which reduces the statistical errors by up to $\sim$20\% compared to the case without weighting. 
}

\begin{document}
\maketitle
\flushbottom

\section{Introduction}\label{sec:intro}

Galaxy-galaxy weak lensing, which can be measured by cross-correlating the positions of lens galaxies with the shapes of background galaxies, is a powerful probe of the galaxy-dark matter connection as well as cosmology \citep[e.g.,][for pioneering works]{2004AJ....127.2544S,2005PhRvD..71d3511S,2006MNRAS.368..715M}. In particular, by combining galaxy-galaxy weak lensing with the auto-correlations of lens galaxies, one can robustly infer cosmological parameters while observationally disentangling the bias uncertainty of the lens galaxies -- the so-called 2$\times$2pt or 3$\times$2pt cosmology \citep{mandelbaum_cosmological_2013,2011PhRvD..83b3008O,2015ApJ...806....2M,2018PhRvD..98d3526A,2021A&A...646A.140H,2022PhRvD.106h3520M,2020JCAP...03..044N,miyatake_hyper_2023,2023PhRvD.108l3521S,2025arXiv250701386Z} \citep[also see Ref.][for a review]{2018ARA&A..56..393M}.

Most galaxy-galaxy weak lensing analyses have been performed using two-point correlation functions in configuration (real) space, with only a few studies using power spectrum in Fourier space \citep{hikage_pseudo-spectrum_2016,2020JCAP...03..044N,faga_dark_2024} \citep[also see Refs.][for studies of cosmic shear power spectrum]{hu_power_2001,brown_shear_2003,hikage_shear_2011,kohlinger_kids-450_2017,hikage_cosmology_2019,alonso_unified_2019,doux_dark_2022,dalal_hyper_2023}. 
Although real- and Fourier-space analyses are theoretically equivalent, this equivalence does not hold exactly in practice. 
Cosmological analyses of weak lensing typically apply scale cuts to mitigate the impact of poorly understood physical processes, such as baryonic effects and other nonlinearities.
The scale cuts break the  equivalence between real- and Fourier-space analyses under the Fourier transformation \citep[e.g. see Appendix~5 in Ref.][for a comparison of cosmology analyses using the real- and Fourier-space cosmic shear two-point correlations]{2020PASJ...72...16H}.
For this reason, it is worthwhile to explore cosmological analyses in both real and Fourier space to extract the  cosmological information and to perform various tests of systematic effects, for ongoing and upcoming galaxy surveys such as Subaru Hyper Suprime-Cam \cite{aihara_hsc_overview} and the Rubin Observatory's LSST\footnote{https://www.lsst.org}.

Therefore, the purpose of this paper is to develop an estimator for measuring the angular power spectrum of galaxy-galaxy weak lensing. To do this, we apply the maximum likelihood method, developed for measuring the power spectra of cosmic microwave background anisotropies \citep{oh_efficient_1999,smith_algorithms_2011,philcox_optimal_2023} and galaxy clustering \citep{philcox_cosmology_2021,philcox_cosmology_2021-1}, to the weak lensing context. 
The estimator allows to correct for survey window effects, such as those arising from survey geometry and bright star masks. We will further account for the spin-2 nature of weak lensing distortions and develop an estimator to simultaneously measure the $E$- and $B$-mode power spectra from galaxy-galaxy weak lensing.
The window-free $E$-mode power spectrum enables straightforward comparison with theoretical predictions in cosmological inference.
To validate our method, we use the weak lensing fields and halo catalogs in the ray-tracing simulations \cite{takahashi_full-sky_2017}, which serve as proxies for source galaxy ellipticities and lens galaxies, respectively, to assess the accuracy of the estimator developed in this paper.

Assuming that lens galaxies in galaxy-galaxy weak lensing are from a spectroscopic sample, where each galaxy has a secure redshift, we also develop a method to estimate the 3D galaxy-matter power spectrum, $P_{\rm gm}(k)$, from the measured galaxy-galaxy weak lensing power spectrum. 
The 3D galaxy-matter power spectrum is one of the fundamental statistical quantities in  galaxy clustering analyses.
We use the same light-cone simulations to test the estimation of $P_{\rm gm}(k)$ and derive an optimal weighting method to reduce  statistical errors in the shot noise limited regime.

We organize this paper as follows. In Section~\ref{sec:g-glensing}, we give a brief overview of the basics of galaxy-galaxy weak lensing. In Section~\ref{sec:quadratic estimator}, we derive a quadratic estimator to measure the $E$- and $B$-mode angular power spectra of galaxy-galaxy weak lensing, following a review of the window-free power spectrum estimator based on the maximum likelihood method.  
In Section~\ref{sec:validation}, we present the main results of this paper: 
we validate our method using the weak lensing fields and halo catalogs in the ray-tracing simulations, and show that the estimator can recover the underlying power spectrum. 
In Section~\ref{sec:application}, we present an application of our method to estimate the 3D galaxy-matter cross-power spectrum, $P_{\rm gm}(k)$, from galaxy-galaxy weak lensing. 
Section~\ref{sec:Conclusion} is devoted to discussion and conclusion.
Throughout this paper, unless otherwise stated, we adopt natural units where the speed of light $c=1$.

\section{Basics of galaxy-galaxy weak lensing}\label{sec:g-glensing}

Galaxy-galaxy weak lensing measurements require two distinct samples of galaxies~\citep{2005MNRAS.361.1287M,2006MNRAS.368..715M}. 
The first sample consists of source (background) galaxies whose shapes are used as tracers of weak lensing effects due to the intervening large-scale structure. The source galaxies are constructed from photometric data,  such as the Subaru HSC data. The second sample consists of lens (foreground) galaxies, which serve as tracers of lensing structures. In this paper we assume that the lens sample is constructed from spectroscopic data, such as the SDSS catalog, and that it has an overlap with the area of
the source galaxy sample.
By cross-correlating the positions of lens galaxies with the shapes of source galaxies, we can measure galaxy-galaxy weak lensing 
signals.

The convergence field for a source galaxy in the angular direction $\B{\theta}$, denoted as $\kappa(\B{\theta})$, is defined as a line-of-sight integral of the three-dimensional matter density fluctuation field, $\delta_{\rm m}(\B{x})$, along the line of sight:
\begin{align}
    \kappa(\B{\theta})=\frac{3}{2}\Omega_{\rm m}H^2_0\int_0^\infty\!\mathrm{d}\chi'~ \chi'g(\chi')a^{-1}(\chi')\delta_{\rm m}(\B{\theta},\chi'), \label{eq:convergence}
\end{align}
where $\chi'$ is a comoving distance and $a(\chi')$ is a scale factor. $g(\chi')$ is a lens efficiency given by
\begin{align}
    g(\chi')=\int_{\chi'}^\infty\!\mathrm{d}\chi_s~ p(\chi_s)\frac{\chi_s-\chi'}{\chi_s} \label{eq:lens efficiency}, 
\end{align}
where $p(\chi_s)$ is the redshift distribution of source galaxies in an average sense, and is normalized such that $\int_0^\infty\!\mathrm{d}z_s~ p(z_s)=1$. 
Note that we assume a flat universe geometry throughout this paper, and the comoving radial distance is given by redshift via the redshift-distance relation, $\chi=\chi(z)$. For the following discussion, we define the inverse of the critical surface mass density, averaged over the redshift distribution of source galaxies, as
\begin{align}
    \Sigma^{-1}_\mr{crit}&(z)\equiv \int_{\chi}^\infty\!\mathrm{d}\chi_s~ p(\chi_s)
    \frac{3H^2_0}{2\rho_{\mr{crit}}}(1+z)\frac{\chi(\chi_s-\chi)}{\chi_s} 
    \equiv
     \frac{3H^2_0}{2\rho_{\mr{crit}}}(1+z)\chi g(\chi),
     \label{eq:Sigma_cr}
\end{align}
where $\rho_{\rm crit}$ is the critical density of the universe today.

The two-dimensional number density fluctuation field of lens galaxies, projected over the redshift interval, is defined as
\begin{align}
\Delta_{\rm g}(\B{\theta})\equiv \int_0^\infty\!\mathrm{d}\chi~ p_{L}(\chi)\delta_{\rm g}(\B{\theta},\chi), \label{eq:projected density}
\end{align}
where $\delta_{\rm g}(\B{x})$ is the three-dimensional number density fluctuation field of galaxies and $p_{L}(\chi)$ denotes the redshift distribution of lens galaxies, normalized as $\int_0^\infty\!\mathrm{d}\chi~ p_{L}(\chi)=1$. 

We now define the angular power spectra for galaxy-galaxy weak lensing. Galaxy shapes, which are direct observables from the data, form a spin-2 field with two components per galaxy. 
A useful method for handling the two components is $E/B$-mode decomposition~\citep[e.g.,][]{2001PhR...340..291B}, in which only the $E$-mode represents the pure weak lensing signal arising from the scalar gravitational potential, while the $B$-mode serves as a diagnostic for systematic effects and/or other nonlinear contributions.
The angular power spectra for the $E/B$ modes of galaxy-galaxy weak lensing are defined as 
\begin{align}
    \ev{\Delta_{\rm g}(\B{\ell})E^*(\B{\ell'})}&=(2\pi)^2\delta^{(2)}_D(\B{\ell}-\B{\ell'})C_{{\rm g}E}(\ell),
    \nonumber \\
    \ev{\Delta_{\rm g}(\B{\ell})B^*(\B{\ell'})}&=(2\pi)^2\delta^{(2)}_D(\B{\ell}-\B{\ell'})C_{{\rm g}B}(\ell),\label{eq:defCell}
\end{align}
where $\delta^{(2)}_D$ is the 2D Dirac delta function. Throughout this paper, we adopt the flat-sky approximation.
Using the Limber approximation \citep{limber}, we can express the $E$-mode angular power spectrum in terms of the underlying 3D power spectrum as
\begin{align}
    C_{{\rm g}E}(\ell)=\frac{3}{2}\Omega_{\rm m}H^2_0\int_0^\infty\! \frac{\mathrm{d}\chi}{\chi}
    p_L(\chi)a^{-1}(\chi)g(\chi)P_{\rm gm}\!\parbra{k=\frac{\ell}{\chi}\ ; z(\chi)}, 
    \label{eq:cl_cge}
\end{align}
where $P_{\rm gm}(k;z)$ is the 3D cross-power spectrum between the matter and galaxy distributions at redshift $z$, and 
redshift is given by the comoving distance via the inverse relation, $z=\chi^{-1}(\chi)$.
Thus, the measured galaxy-galaxy weak lensing signal can be used to estimate $P_{\rm gm}(k)$, as will be discussed in Section~\ref{sec:application}. 
Note that galaxy-galaxy weak lensing is not contaminated by intrinsic alignments, as long as source galaxies are well separated in redshift from lens galaxies.

\section{Quadratic estimator for galaxy-galaxy weak lensing power spectra}
\label{sec:quadratic estimator}

In this section, we derive the quadratic estimator for galaxy-galaxy weak lensing power spectrum. 
An advantage of the quadratic estimator is it allows us to estimate the power spectrum free from survey window effects.

\subsection{Maximum likelihood estimator}\label{subsec:MLE}

We begin by summarizing the basic concept and key features of the quadratic estimator in a simplified
case, following Refs.~\citep{bond_estimating_1998, oh_efficient_1999, philcox_cosmology_2021}.

The derivation of the quadratic estimator starts with a likelihood function that characterizes statistical properties of the observed field. 
For the power spectrum estimator, we assume that the observed field $\B{d}$
(e.g., density fluctuation field) follows a Gaussian likelihood function, given a set of band power parameters, $\B{p}$:
\begin{align}
\mathcal{L}(\B{d}|\B{p})=-2\ln L(\B{d}|\B{p})=\B{d}^TC^{-1}(\B{p})\B{d}+\ln (\mr{det}[C(\B{p})])+\mr{const.}~ ,\label{eq:likelihood}
\end{align}
where $C$ is the covariance matrix of the data vector and $C^{-1}$ is its inverse matrix.
The covariance matrix is 
defined in terms of the underlying 
power spectrum.
\begin{align}
    C(\B{p};\B{x}_i,\B{x}_j)\equiv 
    \langle d(\B{x}_i)d(\B{x}_j)\rangle=W(\B{x}_i) W(\B{x}_j)\int_\B{k} P(k)e^{i\B{k}\cdot (\B{x}_i-\B{x}_j)}+\mathcal{N}(\B{x}_i,\B{x}_j),
     \label{eq:covariance}
\end{align}
where $W(\B{x})$ is the survey window function and $\mathcal{N}$ is a noise contribution such as 
shot noise or shape noise. 
In the simplest case, $W(\B{x})=1$ if $\B{x}$ is inside the survey region, and $W(\B{x})=0$ otherwise. 
In the above equation, we used the collapsed notation as 
$\int_{\B{k}}\equiv (2\pi)^{-n}\int\mathrm{d}^n{\B{k}}$, where 
$n=2$ or 3 for 2D or 3D fields, respectively.

Let us consider a data vector defined on grids and denote the total number of grids as $N_{\mr{grid}}$. 
That is, the data vector is given as $\B{d}=\left\{d(\B{x_1}),\cdots,d(\B{x}_{N_{\mr{grid}}})\right\}$.
The covariance matrix $C$ has dimensions of $N_{\mr{grid}}\times N_{\mr{grid}}$.
This discrete treatment allows us to use a fast Fourier transform (FFT) technique, which makes relevant computations, such as the integrals involved in Fourier transforms, tractable.
Introducing the band power parameters, $\B{p}=\{p_\alpha\}$
we can express the power spectrum as a discrete sum:
$P(k)=\sum_\alpha p_\alpha \Theta_\alpha(|\B{k}|)$, where 
$\Theta_\alpha(|\B{k}|$) is the Heaviside step function; 
$\Theta_\alpha(|\B{k}|)=1$ if $\B{k}$ is inside the $\alpha$-th $k_\alpha$ bin and otherwise 
$\Theta_\alpha(|\B{k}|)=0$. Here we assume that the power spectrum varies slowly enough within each $k_\alpha$ bin.
In this setup, we can compute the partial derivative of the covariance matrix with respect to the $\alpha$-th band power parameter as
\begin{align}
    C_{,\alpha}\equiv \frac{\partial C}{\partial p_\alpha}
    =W(\B{x}_i) W(\B{x}_i)\int_\B{k}\Theta_\alpha(\B{k})e^{i\B{k}\cdot (\B{x}_i-\B{x}_j)} \label{eq:cov_derivative}.
\end{align}

Taylor-expanding the likelihood (Eq.~\ref{eq:likelihood}) around an initial guess of the band power parameters, 
$p_{\alpha,{\rm guess}}$, yields
\begin{align}
\mathcal{L}(\B{p})\simeq \mathcal{L}(\B{p}_{\mr{guess}})+\left.\pdv{\mathcal{L}}{p_\alpha}
\right|_{\B{p}_{\mr{guess}}}(p_{\alpha}-p_{\alpha,\mr{guess}})+
\frac{1}{2}\left.\pdv{\mathcal{L}}{p_\alpha}{p_\beta}\right|_{\B{p}_{\mr{guess}}}
(p_{\alpha}-p_{\alpha,\mr{guess}})(p_{\beta}-p_{\beta,\mr{guess}})~,
\label{eq:likelihood_expansion}
\end{align}
where we have neglected the higher-order terms assuming that the difference between $\{p_{\alpha,\mr{guess}}\}$ and the true power spectrum $\{p_{\alpha,\mr{true}}\}$ is sufficiently small.
The true band powers $\{p_{\alpha,\mr{true}}\}$ can be estimated by minimizing Eq.~(\ref{eq:likelihood_expansion}):
\begin{align}
    p_{\alpha,\mr{true}}&\simeq p_{\alpha,\mr{guess}}-\frac{1}{2}F^{-1}_{\alpha\beta}\left.\pdv{\mathcal{L}}{p^\beta}\right|_{\B{p}_{\mr{guess}}},
    \label{eq:MLEpre}
\end{align}
where $F_{\alpha\beta}$ is the Fisher matrix defined by
\begin{align}
F_{\alpha\beta}\equiv\frac{1}{2}\left.\pdv{\mathcal{L}}{p^\alpha}{p^\beta}\right|_{\B{p}_{\mr{guess}}},
\label{eq:Fisher_def}
\end{align}
and $F^{-1}$ is the inverse of the Fisher matrix. 
Computing the first and second derivatives of the log likelihood in Eq.~(\ref{eq:likelihood_expansion}) using Eq.~(\ref{eq:likelihood}) yields an estimator of the $\alpha$-th band power parameter:
\begin{align}
    \hat{p}_\alpha&=p_{\alpha,\mr{guess}}+\frac{1}{2}\sum_\beta F^{-1}_{\alpha\beta}\mr{Tr}\sqbra{C^{-1}_{\mr{guess}}C_{,\beta}C^{-1}_{\mr{guess}}(\B{dd^T}-C_{\mr{guess}})},
        \label{eq:MLEformula}
\end{align}
with
\begin{align}
    F_{\alpha\beta}&=\frac{1}{2}\mr{Tr}[C^{-1}_{\mr{guess}}C_{,\alpha}C^{-1}_{\mr{guess}}C_{,\beta}],
\end{align}
where $C_{\rm guess}$ denotes the covariance matrix evaluated using the initial guess of the 
power spectrum. 
Here the covariance $C_{\rm guess}$ includes contributions from observational effects such as shot noise and shape noise. As discussed in Ref.~\cite{philcox_cosmology_2021}, the inverse of covariance matrix in the estimator acts as a weighting of the data vector, e.g., the FKP weight \cite{feldman_power_1994} used in galaxy clustering analyses. 
If the guessed power spectrum and covariance are close to the underlying true ones, the above estimator provides an optimal estimate of the band power in each $k$ bin.

We can easily check that Eq.~(\ref{eq:MLEformula}) is an unbiased estimator of the true band power in an average sense.
Because $C=\sum_\alpha C_{,\alpha}p_\alpha+\mathcal{N}$ and $\ev{\B{d}\B{d}^T}-C_{\mr{guess}}=
\sum_\mu C_{,\mu}
(p_{\mu,\mr{true}}-p_{\mu, \mr{guess}})$, we can find that the ensemble average 
of Eq.~(\ref{eq:MLEformula}) yields $\langle \hat{p}_\alpha\rangle=p_{\alpha,\mr{true}}$.

The estimator, given by Eq.~(\ref{eq:MLEformula}), can be generalized by using an arbitrary weight
matrix $H$ in place of the initial-guess covariance matrix $C_{\rm guess}$:
\begin{align}
    \hat{p}_\alpha&=p_{\alpha,\mr{guess}}+\frac{1}{2}\sum_\beta F^{-1}_{\alpha\beta}\mr{Tr}\sqbra{H^{-1}C_{,\beta}H^{-1}
    (\B{d}\B{d}^T-C_{\rm guess})}\nonumber \\
    &=\frac{1}{2}\sum_\beta F^{-1}_{\alpha\beta}\mr{Tr}\sqbra{H^{-1}C_{,\beta}H^{-1}
    (\B{d}\B{d}^T-\mathcal{N})},
        \label{eq:estimator_general}
\end{align}
with
\begin{align}
    F_{\alpha\beta}&=\frac{1}{2}\mr{Tr}[H^{-1}C_{,\alpha}H^{-1}C_{,\beta}]~.
\end{align}
As we show in Appendix \ref{subsec:correspondence}, this estimator becomes equivalent to the pseudo-$C_\ell$ estimator, when we take $H\propto I$, where $I$ is the identity matrix.

\subsection{Quadratic estimator for
$C_{{\rm g}E}(\ell)$ and $C_{{\rm g}B}(\ell)$}
\label{subsec:gglensing}

In this section, we describe an estimator for galaxy-galaxy weak lensing. As described in Section~\ref{sec:g-glensing}, galaxy-galaxy weak lensing measurements require two different samples of galaxies: the one of
source galaxy shapes, which have two degrees of freedom, and the other of lens galaxies.
Therefore, a naive application of the maximum likelihood estimator to galaxy-galaxy weak lensing requires that the data vector in the likelihood function includes all three fields, which requires to include the covariance for all the three fields.
Rather than pursuing this direction, in this paper we explore a closed-form estimator only for the power spectrum of galaxy-galaxy weak lensing.

The data vector including all three fields is defined as
\begin{align}
\B{d}&\equiv \left\{\B{\Delta}_{\rm g},\B{\gamma}_1,\B{\gamma}_2\right\}\nonumber\\
&=
\left\{\Delta_{\rm g}(\B{\theta}_1),\dots,\Delta_{\rm g}(\B{\theta}_{N_{\rm grid}}),
\gamma_1(\B{\theta}_1),\dots,\gamma_1(\B{\theta}_{N_{\rm grid}}),\gamma_2(\B{\theta}_1),\dots,\gamma_2(\B{\theta}_{N_{\rm grid}})
\right\}.
\end{align}
Therefore, the data vector $\B{d}$ has $3N_{\rm grid}$ elements. 
We use the bold notation such as $\B{\Delta}_{\rm g}$ to denote the data vector that has a dimension of $N_{\rm grid}$; 
that is, e.g., $\B{\Delta}_{\rm g}=\{\Delta_{\rm g}(\B{\theta}_1),\dots,\Delta_{\rm g}(\B{\theta}_{N_{\rm grid}})\}$.
The cross-correlation functions between the galaxy number density field and the shear fields are given as
\begin{align}
C_{\gamma_1{\rm g}}(\B{\theta}_1,\B{\theta}_2)
\equiv W_\gamma(\B{\theta}_1) W_{\rm g}(\B{\theta}_2)
\int_{\B{\ell}}~e^{i\B{\ell}\cdot(\B{\theta}_1-\B{\theta}_2)}
\left[P_{{\rm g}E}(\ell)\cos2\phi_{\B{\ell}}-P_{{\rm g}B}(\ell)\sin2\phi_{\B{\ell}}\right], \nonumber\\
C_{\gamma_2{\rm g}}(\B{\theta}_1,\B{\theta}_2)
\equiv W_\gamma(\B{\theta}_1) W_{\rm g}(\B{\theta}_2)
\int_{\B{\ell}}~e^{i\B{\ell}\cdot(\B{\theta}_1-\B{\theta}_2)}
\left[P_{{\rm g}E}(\ell)\sin2\phi_{\B{\ell}}+P_{{\rm g}B}(\ell)\cos2\phi_{\B{\ell}}\right], 
\end{align}
Note that the order of the subscript ``$\gamma{\rm g}$'' and the argument ``$(\B{\theta}_1,\B{\theta}_2)$'' is meaningful;
e.g., generally $C_{\gamma_1{\rm g}}(\B{\theta}_1,\B{\theta}_2)\ne C_{{\rm g}\gamma_1}(\B{\theta}_1,\B{\theta}_2)$, when the window functions for the galaxy and shear fields are different.

Taking only the diagonal elements 
of the weight matrix $H^{-1}$ in the estimator (Eq.~\ref{eq:estimator_general}), 
we can derive a quadratic estimator of the band powers in the $\alpha$-th $\ell$ bin for the $E$- and $B$-mode cross power spectra, $C_{{\rm g}E}$ and $C_{{\rm g}B}$ in Eq.~(\ref{eq:defCell}), as
\begin{align}
\left(
\begin{array}{ll}
\hat{p}_{{\rm g}E,\alpha}\\
\hat{p}_{{\rm g}B,\alpha}
\end{array}
\right)
=
\left(
\begin{array}{ll}
(F^{-1})^{EE}_{\alpha\beta} & (F^{-1})^{EB}_{\alpha\beta} \\
(F^{-1})^{BE}_{\alpha\beta} & (F^{-1})^{BB}_{\alpha\beta}
\end{array}
\right)
\left(
\begin{array}{ll}
\hat{q}_{{\rm g}E,\beta}\\
\hat{q}_{{\rm g}B,\beta}
\end{array}
\right), 
\label{eq:gglensing_quadratic_estimator}
\end{align}
where
\begin{align}
\hat{q}_{{\rm g}E,\alpha}&=(H^{-1}_\gamma \B{\gamma}_1)^T
C_{\gamma{\rm g},\alpha}^{c}(H^{-1}_{\rm g}\B{\Delta}_{\rm g}) +(H^{-1}_\gamma\B{\gamma}_2)^T 
C_{\gamma{\rm g},\alpha}^{s}(H^{-1}_{\rm g}\B{\Delta}_{\rm g}), \nonumber\\
    \hat{q}_{{\rm g}B,\alpha}&=-(H^{-1}_\gamma\B{\gamma}_1)^T C_{\gamma{\rm g},\alpha}^{s}(H^{-1}_{\rm g}\B{\Delta}_{\rm g})
     +(H^{-1}_\gamma\B{\gamma}_2)^T C_{\gamma{\rm g},\alpha}^{c}(H^{-1}_{\rm g}\B{\Delta}_{\rm g}),
     \label{eq:gglensing_estimator_q}
\end{align}
$H^{-1}_\gamma$ or $H^{-1}_{\rm g}$ is the weight for the shear or galaxy field, respectively, and
the Fisher matrix, $F=\begin{pmatrix}F^{EE} & F^{EB}\\ F^{BE} & F^{BB}\end{pmatrix}$, has components defined as
\begin{align}
    {F}_{\alpha\beta}^{EE}
    &=F_{\alpha\beta}^{BB}=\mr{Tr}\sqbra{H_{\gamma}^{-1}C^{c}_{\gamma{\rm g},\alpha}
    \ H^{-1}_{\rm g} C^{c}_{{\rm g}\gamma,\beta}}+\mr{Tr}\sqbra{H^{-1}_{\gamma} C^{s}_{\gamma{\rm g},\alpha}\ H^{-1}_{\rm g} 
    C^{s}_{{\rm g}\gamma,\beta}}, \nonumber\\
    {F}_{\alpha\beta}^{EB}
    &=-F_{\alpha\beta}^{BE}=\mr{Tr}\sqbra{H^{-1}_\gamma C^{s}_{\gamma{\rm g},\alpha}\ H^{-1}_{\rm g} C^{c}_{{\rm g}\gamma,\beta}}-\mr{Tr}\sqbra{H^{-1}_\gamma C^{c}_{\gamma{\rm g},\alpha}\ H^{-1}_{\rm g} C^{s}_{{\rm g}\gamma,\beta}}.
    \label{eq:gg_lensing_estimator_fisher}
\end{align}
with 
\begin{align}
C^u_{\gamma{\rm g},\alpha}(\B{\theta}_1,\B{\theta}_2)\equiv 
W_\gamma(\B{\theta}_1)W_{\rm g}(\B{\theta}_2)
\int_{\B{\ell}}e^{i\B{\ell}\cdot(\B{\theta}_1-\B{\theta}_2)}f_u(2\phi_{\B{\ell}})
\Theta_{\alpha}(\B{\ell}),
\end{align}
and $f_u(2\phi_{\B{\ell}})=\cos2\phi_{\B{\ell}}$ or $\sin2\phi_{\B{\ell}}$ for $u=c$ or $s$, respectively. 
The band powers, $\hat{p}_{{\rm g}E}$ and $\hat{p}_{{\rm g}B}$, are estimated from the quadratic products of the galaxy and shear fields, $\B{\Delta}_{\rm g}$ and $\B{\gamma}_i$. Here, we ignored the noise term $\mathcal{N}$ in Eq.~(\ref{eq:estimator_general}), because the shot noise vanishes for the cross-correlation.
The above estimator is in closed form, meaning that the galaxy-galaxy weak lensing power spectra can be estimated without requiring knowledge of the auto-power spectra of $\B{\Delta}_{\rm g}$ and~$\B{\gamma}_i$.

A direct computation of the Fisher matrices is computationally expensive. Instead, as shown in Refs.~\cite{oh_efficient_1999,smith_algorithms_2011,philcox_cosmology_2021,philcox_cosmology_2021-1}, the Monte Carlo method can be used. By introducing ``ancillary'' vector fields, denoted as $\{\B{a}\}$, one could compute the trace required for the Fisher matrix calculation as
\begin{align}
\mr{Tr}\sqbra{H^{-1}_{\gamma}
    C^{c}_{\gamma{\rm g},\alpha}H^{-1}_{{\rm g}}C^{c}_{{\rm g}\gamma,\beta}}&=
    \left\langle H^{-1}_\gamma C^{c}_{\gamma{\rm g},\alpha}H^{-1}_{{\rm g}}C^{c}_{{\rm g}\gamma,\beta}
    A^{-1}\B{a}\B{a}^T
    \right\rangle \nonumber\\
    &= \left\langle ( C^{c}_{\gamma{\rm g},\alpha}H^{-1}_{\gamma}\B{a})^T
    H^{-1}_{{\rm g}}(C^{c}_{{\rm g}\gamma,\beta}
    A^{-1}\B{a})
    \right\rangle,
    \label{eq:MonteCarlo_method}
\end{align}
where the notation $\langle\hspace{0.5em}\rangle$ denotes the ensemble average, which is practically computed using a sufficient number of realizations of the simulated $\{\B{a}\}$ maps, and $A=\langle \B{a}\B{a}^T\rangle$. The quantities in parentheses on the 2nd line of the r.h.s. are vector quantities of $N_{\rm grid}$ dimension, and the trace can be obtained by taking the dot product of the two vectors.

In summary, we use the estimator, given by Eqs.~(\ref{eq:gglensing_quadratic_estimator})--(\ref{eq:gg_lensing_estimator_fisher}), for estimating the band powers of $C_{{\rm g}E}$ and $C_{{\rm g}B}$ power spectra, using the Monte Carlo method such as that given by Eq.~(\ref{eq:MonteCarlo_method}).

\subsection{FFT implementation}
\label{subsec:FFTimplement}

A naive implementation of Eqs.~(\ref{eq:gglensing_quadratic_estimator})--(\ref{eq:gg_lensing_estimator_fisher}) requires a computation with a cost of $O(N_{\rm grid}^2)$. Since the dimension of $N_{\rm grid}$ is high, typically $N_{\rm grid}\sim 10^6$, the computation can be expensive. 
Instead, we use the fast Fourier transform (FFT) method~\citep{philcox_cosmology_2021,philcox_cosmology_2021-1} to efficiently perform these computations, as described below.

To use the Fourier transform, we first assign the galaxy shape and galaxy number density fields to grids and then use the discrete summation. The terms in $\hat{q}_{E,\alpha} $ and $\hat{q}_{B,\alpha}$ (Eq.~\ref{eq:gglensing_estimator_q}) can be computed as
\begin{align}
(H^{-1}_\gamma \B{\gamma}_i)^T
C_{\gamma{\rm g},\alpha}^{u}(H^{-1}_{\rm g}\B{\Delta}_{\rm g})
&=\sum_{\B{\theta}_1,\B{\theta}_2}
\left[H^{-1}_\gamma\B{\gamma}_i\right]\!(\B{\theta}_1)
C^{u}_{\gamma{\rm g},\alpha}(\B{\theta}_1,\B{\theta}_2)
\left[H^{-1}_{\rm g}\B{\Delta}_{\rm g}\right]\!(\B{\theta}_2)\nonumber\\
&\hspace{-8em}= \sum_{\B{\theta}_1,\B{\theta}_2}\int_{\B{\ell}}\! \left[H^{-1}_\gamma\B{\gamma}_i\right]\!(\B{\theta}_1)W_\gamma(\B{\theta}_1)
\Theta_{\alpha}(\B{\ell})f_u(2\phi_\B{\ell})e^{i\B{\ell}\cdot(\B{\theta}_1-\B{\theta}_2)}W_{\mr{g}}(\B{\theta}_2)\left[H^{-1}_{\rm g}\B{\Delta}_{\rm g}\right]\!(\B{\theta}_2)
\nonumber\\
&\hspace{-8em}=\int_{\B{\ell}}\! \Theta_{\alpha}(\B{\ell})f_u(2\phi_\B{\ell})
{\cal F}^*\!\Big[\!\left(H^{-1}_\gamma\B{\gamma}_i\right)\!(\B{\theta})~ W_\gamma(\B{\theta})\Big]\!(\B{\ell})
{\cal F}\Big[\!\left(H^{-1}_{\rm g}\B{\Delta}_{\rm g}\right)(\B{\theta})~ W_{\mr{g}}(\B{\theta})\Big]\!(\B{\ell}),
\label{eq:calc_first term}
\end{align}
where ${\cal F}$ denotes the Fourier transform, given by
\begin{align}
{\cal F}\!\left[\left(H^{-1}_{{\rm g}}\B{\Delta}_{\rm g}\right)\!(\B{\theta})~ W_{\rm g}(\B{\theta})\right]\!(\B{\ell})\simeq 
\int_{\B{\theta}} W_{\rm g}(\B{\theta})\left(H^{-1}_{{\rm g}}\B{\Delta}_{\rm g}\right)\!(\B{\theta})
e^{-i\B{\ell}\cdot\B{\theta}}.\label{eq:fourier transform}
\end{align}
Since the data vector is given on discritized $N_{\rm grid}$ grids, we can use the FFT to perform the above computation.
Note that we consider a diagonal matrix for the weight $H^{-1}$, so computations such as $[H^{-1}_{\rm g}\B{\Delta}_{\rm g}]$ are inexpensive  and can be performed on a per-grid basis. Here we assume that the data vector is given over a rectangular-shaped region, e.g., using the zero-padding method to enclose an arbitrary survey geometry; the window function $W(\B{x})$ accounts for both masked and zero-padded regions. 

Next let us consider the Fisher matrix. As we described below Eq.~(\ref{eq:MonteCarlo_method}), we can compute the Fisher matrix using a realization of the ancillary field $\B{a}$ for which we assume to be a Gaussian field.
For $A=\langle \B{a}\B{a}^T\rangle$, we adopt $A=H_\gamma$, i.e. the inverse of the diagonal weight matrix for the shear field, $H_{\gamma}^{-1}$. In this case, the term that is involved in the Fisher matrix computation can be computed using the FFT method as 
\begin{align}
\mr{Tr}\sqbra{H^{-1}_\gamma C^{u}_{\gamma{\rm g},\alpha}H^{-1}_{\rm g}C^{v}_{{\rm g}\gamma,\beta}A^{-1}\B{a}\B{a}^T}
\nonumber\\
&\hspace{-10em}=\sum_{\B{\theta}_1,\B{\theta}_2,\B{\theta}_3} 
(H_\gamma^{-1}\B{a})(\B{\theta}_1)C^{u}_{\gamma{\rm g},\alpha}(\B{\theta}_1,\B{\theta}_2)
H_{\rm g}^{-1}(\B{\theta}_2)C^{v}_{{\rm g}\gamma,\beta}(\B{\theta}_2,\B{\theta}_3)(H^{-1}_\gamma\B{a})(\B{\theta}_3)\nonumber\\
&\hspace{-10em}=\sum_{\B{\theta}_1,\B{\theta}_2,\B{\theta}_3} \int_{\B{\ell}}\int_{\B{\ell'}}
\sqbra{(W_\gamma H_\gamma^{-1}\B{a})}(\B{\theta}_1)\Theta_{\alpha}(\B{\ell})f_u(2\phi_\B{\ell})e^{i\B{\ell}\cdot(\B{\theta}_1-\B{\theta}_2)}
\sqbra{H_{\rm g}^{-1}(\B{\theta}_2)W_{\rm g}(\B{\theta}_2)^2}
\nonumber\\
&\hspace{0em}\times 
\Theta_{\beta}(\B{\ell}')f_v(2\phi_{\B{\ell}'})e^{i\B{\ell}'\cdot(\B{\theta}_2-\B{\theta}_3)}\sqbra{(W_\gamma
H^{-1}_\gamma\B{a})}(\B{\theta}_3)\nonumber\\
&\hspace{-10em}= \sum_{\B{\theta}_2}H_{\rm g}^{-1}(\B{\theta}_2)W_{\rm g}(\B{\theta}_2)^2\int_{\B{\ell}}\Theta_{\alpha}(\B{\ell})f_u(2\phi_\B{\ell})\mathcal{F}^*\sqbra{(W_\gamma H_\gamma^{-1}\B{a})}(\B{\ell})e^{-i\B{\ell}\cdot\B{\theta}_2}\nonumber\\
&\times  \int_{\B{\ell}'}\Theta_{\beta}(\B{\ell}')f_v(2\phi_{\B{\ell}'})
\mathcal{F}\sqbra{(W_\gamma H_\gamma^{-1}\B{a})}(\B{\ell}')e^{i\B{\ell}'\cdot\B{\theta}_2}\nonumber\\
&\hspace{-10em}= \sum_{\B{\theta}_2}H_{\rm g}^{-1}(\B{\theta}_2)W_{\rm g}(\B{\theta}_2)^2 \mathcal{F}^{-1*}\Big[\Theta_{\alpha}(\B{\ell})f_u(2\phi_\B{\ell})\mathcal{F}^*\sqbra{(W_\gamma H_\gamma^{-1}\B{a})}(\B{\ell})\Big](\B{\theta}_2)\nonumber\\
&\times \mathcal{F}^{-1}\Big[\Theta_{\beta}(\B{\ell}')f_v(2\phi_{\B{\ell}'})\mathcal{F}
\sqbra{(W_\gamma H^{-1}_\gamma\B{a})}(\B{\ell}')\Big](\B{\theta}_2),
\label{eq:fisher_MonteCarlo_fft}
\end{align}
where ${\cal F}^{-1}$ denotes the inverse Fourier transform, compared to Eq.~(\ref{eq:fourier transform}). 
The computational cost for the above term is $O(N_{\rm grid}\log N_{\rm grid})$ for each realization of the simulated $\B{a}$ map. A computation of each of the Fisher matrices requires to use a sufficient number of the realizations of $\B{a}$. In this paper, we will use $O(100)$ realizations. 
Thus, by using the FFT method, we can implement the quadratic estimator for the band powers of the galaxy-galaxy weak lensing 
using  grids with $N_{\rm grid}\sim O(10^6)$ and $O(100)$ realizations of $\{\B{a}\}$.
By taking $A=H_\gamma$, we can make $F_{\alpha\beta}^{EE}$ a symmetric matrix for each realization of $\B{a}$; 
that is, $F^{EE}_{\alpha\beta}=F^{EE}_{\beta\alpha}$. Similarly we have $F^{EB}_{\alpha\beta}=-F^{EB}_{\beta\alpha}$. 
However, we note that, even if we take the simplest choice of $A=I$, the results do not largely change compared to 
the results we will show below.

For a trivial survey geometry, such as a square or rectangular shape, and in the absence of masked region, i.e. $W_{\rm g}=W_\gamma=1$ on all grids and assuming $H^{-1}=I$, we can find that the calculation, such as that given by Eq.~(\ref{eq:fisher_MonteCarlo_fft}), leads to $F_{\alpha\beta}^{EE}\propto \delta^K_{\alpha\beta}$ due to the facts that $\langle\B{a}\B{a}^T\rangle=I$ and the orthogonality of the Fourier modes for a finite area, $\int_{\B{\theta}}e^{i(\B{\ell}-\B{\ell}')\cdot\B{\theta}}\propto \delta^K_{\B{\ell}\B{\ell}'}$, where $\delta^K_{ij}$ is the Kronecker delta function. In this case, we can also find $F^{EB}_{\alpha\beta}=0$.
Conversely, the survey window effect induces non-zero off-diagonal elements in the Fisher matrices.
Non-zero $F_{\rm \alpha\beta}^{EB}$ describes the leakage of the $E$ (and generally $B$) mode into the $B$ (or $E)$ mode due to the survey window effect.

\section{Validation of the quadratic estimator}
\label{sec:validation}

\subsection{Ray-tracing simulations} 
\label{subsec:simulation_maps}

\begin{figure}[h]
\centering 
\includegraphics[width=1\textwidth]{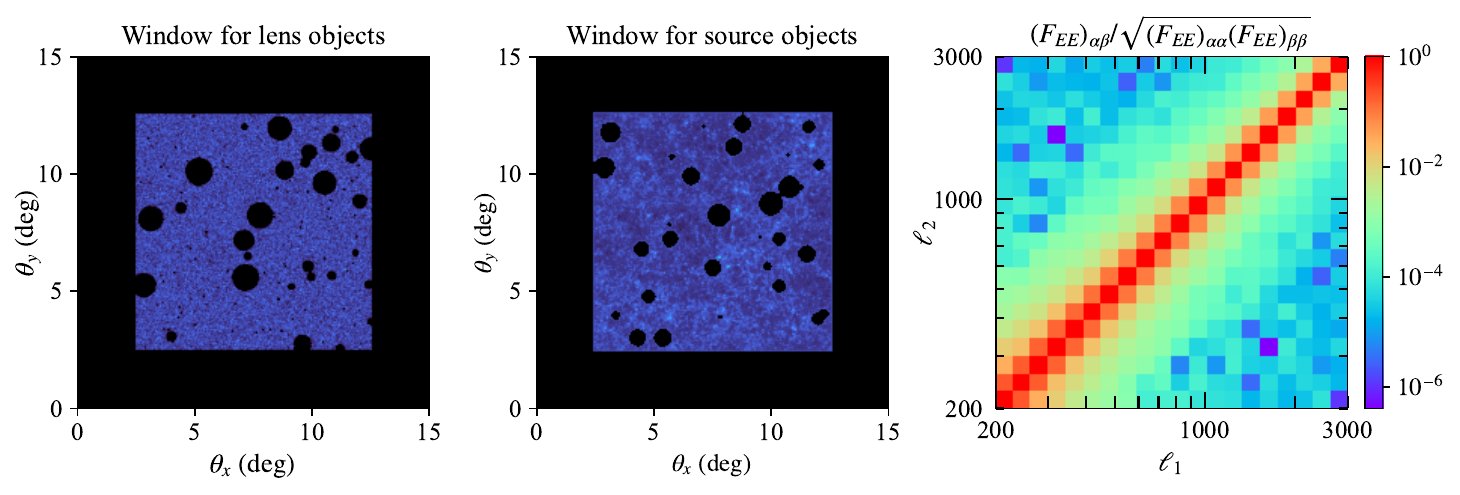}
\vspace{-0.5cm}
\caption{{\it Left or middle panel}: The survey window for the lens or source sample, respectively, which we consider in this paper as a working example.
The region for the lens (halos) or source galaxies (lens galaxies at $0.3<z_L<0.5$ and weak lensing shear at $z_s=1.03$) is defined 
  as a rectangular shape with $\ang{10}\times\ang{10}$ area, which is extracted from the full-sky 
 ray-tracing simulations (see text for details).
 The color denotes the distribution of random points 
 for the lens plane and the lensing convergence ($\kappa$) for the source plane, respectively.
 We introduce randomly placed circular masked regions, for example to mimic bright stars, which cover about 15\% 
 of the data region. Here, we assume that each masked region has a random radius and a random location, and we consider independent mask configurations for the lens and source samples.
Furthermore, we add zero-padded regions surrounding the data region and use a total area of $\ang{15}\times \ang{15}$ area to apply the quadratic estimator with the Fourier transformation method.  
{\it{Right panel}}: The Fisher matrix $F^{EE}_{\alpha\beta}$ (see Eq.~\ref{eq:gg_lensing_estimator_fisher}) computed for the survey windows shown here. We use the Monte Carlo method to estimate $F^{EE}_{\alpha\beta}$ using 300 realizations of the ancillary field $\{\B{a}\}$. Here we consider 20 logarithmically spaced bins in the multipole range $200\le \ell\le 3000$ and show the normalized correlation coefficients, defined as $c_{\alpha\beta}\equiv F^{EE}_{\alpha\beta}/[F^{EE}_{\alpha\alpha}F^{EE}_{\beta\beta}]^{1/2}$, for illustrative purposes. 
For the ideal case with no window effect, we have $c_{\alpha\beta}=\delta^K_{\alpha\beta}$.
}
\label{fig:window_fisher}
\end{figure}
We use the full-sky ray-tracing simulations constructed in Ref.~\citep{takahashi_full-sky_2017} to validate our method. 
Ref.~\citep{takahashi_full-sky_2017} performed a set of $N$-body simulations with $2048^3$ particles in cosmological volumes and used them to construct lensing fields and halo catalogues in the light cone volume. 
They employed the $\Lambda$CDM model that is consistent with the WMAP data \cite{hinshaw_nine-year_2013}; to be more precise, the $\Lambda$CDM model is given by $\Omega_{\rm c}=0.233, \Omega_{\rm b}=0.046$, $\Omega_\Lambda=0.721, h=0.7, n_s=0.97$, and $\sigma_8=0.82$. 
Each of the 108 light-cone simulation realizations contains halo catalogs as well as convergence and shear fields at multiple redshifts, given in the \texttt{HEALPix} pixelizations\footnote{The full-sky simulations are available from this \href{http://cosmo.phys.hirosaki-u.ac.jp/takahasi/allsky_raytracing/}{URL}.}. 
The simulated shear field at a given redshift and sky position represents the weak lensing distortion effect on a galaxy located at the redshift and the angular position. 
In this paper we use the simulations provided on \texttt{HEALPix} pixels with $N_{\rm side}=8192$, which corresponds to pixel size of $0.43$~arcmin.
Therefore, the simulated shear field is reliable up to $\ell\sim 4000$, and we thus consider the power spectrum only up to this multipole in this paper.
We also use halos in each realization as proxies for lens galaxies.

We extract 9 non-overlapping regions from each of the 108 realizations. 
Each extracted region covers an area of $\ang{10} \times \ang{10}$, yielding a total of 972$(=108\times 9)$ independent realizations.
Throughput this paper, we adopt the flat-sky approximation taking into account the basis transformation of the shear field
(Terawaki et al. in preparation). In this process, the simulation data defined in the \texttt{HEALPix} spherical coordinates are projected onto the 2D flat coordinates using the Nearest Grid Point (NGP) method \citep{sefusatti_accurate_2016}.
We then consider a $\ang{15}\times\ang{15}$ region for performing the FFT, employing zero-padding around the data region in each realization, as shown in the {\it{left/middle}} panels of Fig.~\ref{fig:window_fisher}.
The zero-padded region can improve the accuracy of the correction for the survey window effect.
We adopt $1950^2$ grids for each region, with each grid having a size of 0.46~arcmin, which is comparable with the original \texttt{HEALPix} pixel size.

For the weak lensing data, we use the simulated weak-lensing fields, $\kappa$, $\gamma_1$ and $\gamma_2$ at a single source redshift, $z_s=1.03$, in each realization, where $\kappa$ is the convergence field and $\gamma_1$ and $\gamma_2$ are the shear fields.
In this paper we use ``distortion'' \cite{2002AJ....123..583B} as a proxy of observed galaxy ellipticities, where the distortion amplitude for an ellipse-shaped galaxy image is given by $(a^2-b^2)/(a^2+b^2)$, with $a$ and $b$ denoting the major and minor axes, respectively.
We generate intrinsic shape components on each grid assuming a Gaussian distribution with a standard deviation of $\sigma_\epsilon=0.2$ for each component. We then simulate the ``observed'' distortion field on each grid by adding the intrinsic shapes to the simulated lensing fields, following Eqs.~(24) and (25) in Ref.~\cite{2019MNRAS.486...52S}.
Since the intrinsic rms shape per component for Subaru HSC-type data is $e_{\rm rms}\simeq 0.4$ (see Table~1 in Ref.~\cite{hikage_cosmology_2019}) \citep[also see][]{mandelbaum_cosmological_2013} and the grid area of our simulated map is 0.21~arcmin$^2$, the assumed value of $\sigma_\epsilon=0.2$ corresponds to $\bar{n}_s\simeq 19~$arcmin$^{-2}$ for the average number density of source galaxies in each grid; $0.2\simeq 0.4/[19\times 0.21]^{1/2}$. 
This is comparable with the number density 
for an HSC-type survey, as can be found from Table~1 of Ref.~\cite{hikage_cosmology_2019}.
Since we simulate the distortion field, we include the ``responsivity'' \cite{2002AJ....123..583B} to estimate the shear field from the simulated ellipticities:
\begin{align}
    \gamma_i(\B{\theta})= w(\B{\theta})\frac{{\epsilon}_i^{\rm{obs}}(\B{\theta})}{2\mathcal{R}}, \label{eq:shear_estimation}
\end{align}
where $\mathcal{R}$ is the responsivity defined as $\mathcal{R}\equiv1-\ev{e^2_{\mr{rms}}}_w$ where $\ev{\hspace{1em}}_w$ denotes the weighted average over the grids used in the statistics. 
For the weight we adopt $w(\B{\theta})=1/[e^2_{\mr{rms}}(\B{\theta})+\sigma^2_e(\B{\theta})]$, where $e_{\rm rms}$ is the simulated intrinsic ellipticities per component on each grid and $\sigma_e$ is the measurement error. 
For simplicity, we assume a constant value of $\sigma_e=0.1$ in this paper.

For the lens galaxies, we use halos obtained from the same light-cone simulations~\cite{takahashi_full-sky_2017}.
We use halos with $M>10^{13}~h^{-1}M_\odot$ in the redshift range $0.3<z<0.5$, which correspond to the typical host-halo masses of BOSS galaxies \citep{miyatake_hyper_2023}. 
We use the random catalog, $n_{\rm r}(\B{\theta})$, generated using random points 50 times the number of halos in the 2D (RA and dec) plane. 
Note that, since we consider only the projected number density field of halos in this paper, we need not to incorporate a redshift distribution for the random catalog to mimic that of the halos.
Then, the density fluctuation field of halos
on each grid is calculated as $d_{\rm g}(\B{\theta})=n_{\rm g}(\B{\theta})-\alpha n_{\rm r}(\B{\theta})$, where $\alpha=1/50$.

Then, we generate masked regions such as those due to bright masks in the simulated maps.
For generality, we consider two different configurations of masked regions for the lens field
and the shear field, as shown in the {\it{left}} and {\it{middle}} panels of Fig.~\ref{fig:window_fisher}. For each field, we generate circular masked regions, assigning a randomly chosen radius to each circle. Note that we define the survey window using $W(\B{\theta})=0$ or 1 on each grid, meaning $W(\B{\theta})=0$ is assigned to any grid that is intersected by 
a masked circle.
The fraction of the masked area relative to the data region is assumed to be about 15\%, which is typical of actual weak lensing data \citep[e.g., see Fig.~2 in Ref.][]{li_three-year_2022}.

For the power spectrum estimation from the simulated maps, we employ 20 logarithmically spaced bins over the multipole range $200\le \ell\le 3000$. 
The {\it right} panel of Fig.~\ref{fig:window_fisher} shows the $20\times 20$ elements of the Fisher matrix, $F^{EE}_{\alpha\beta}$ (Eq.~\ref{eq:gg_lensing_estimator_fisher}), computed using 300 realizations of the ancillary $\B{a}$ field (see Eq.~\ref{eq:MonteCarlo_method}). 
For illustrative purposes, the figure shows the normalized correlation coefficients, defined as $F^{EE}_{\alpha\beta}/[F^{EE}_{\alpha\alpha}F^{EE}_{\beta\beta}]^{1/2}$, so that the diagonal elements are unity by definition and non-zero off-diagonal elements indicate the strength of cross-correlations between different multipole bins. 
We again note that, for the case of a trivial survey geometry with no masked regions, $F^{EE}_{\alpha\beta}\propto \delta^K_{\alpha\beta}$ as discussed below Eq.~(\ref{eq:fisher_MonteCarlo_fft}). 
The figure clearly shows that the Fisher matrix exhibits non-zero cross-correlations between the neighboring bins, due to the survey window effect. 
The quadratic estimator examined in this paper is designed to correct for this survey window effect, 
yielding a window-free estimate of the underlying power spectrum of galaxy-galaxy weak lensing.

\subsection{Results: the accuracy of the quadratic estimator for $C_{{\rm g}E}(\ell)$}
\label{subsec:results_cell_reconstruction}

\begin{figure}[t]
\centering 
\includegraphics[width=0.95\textwidth]{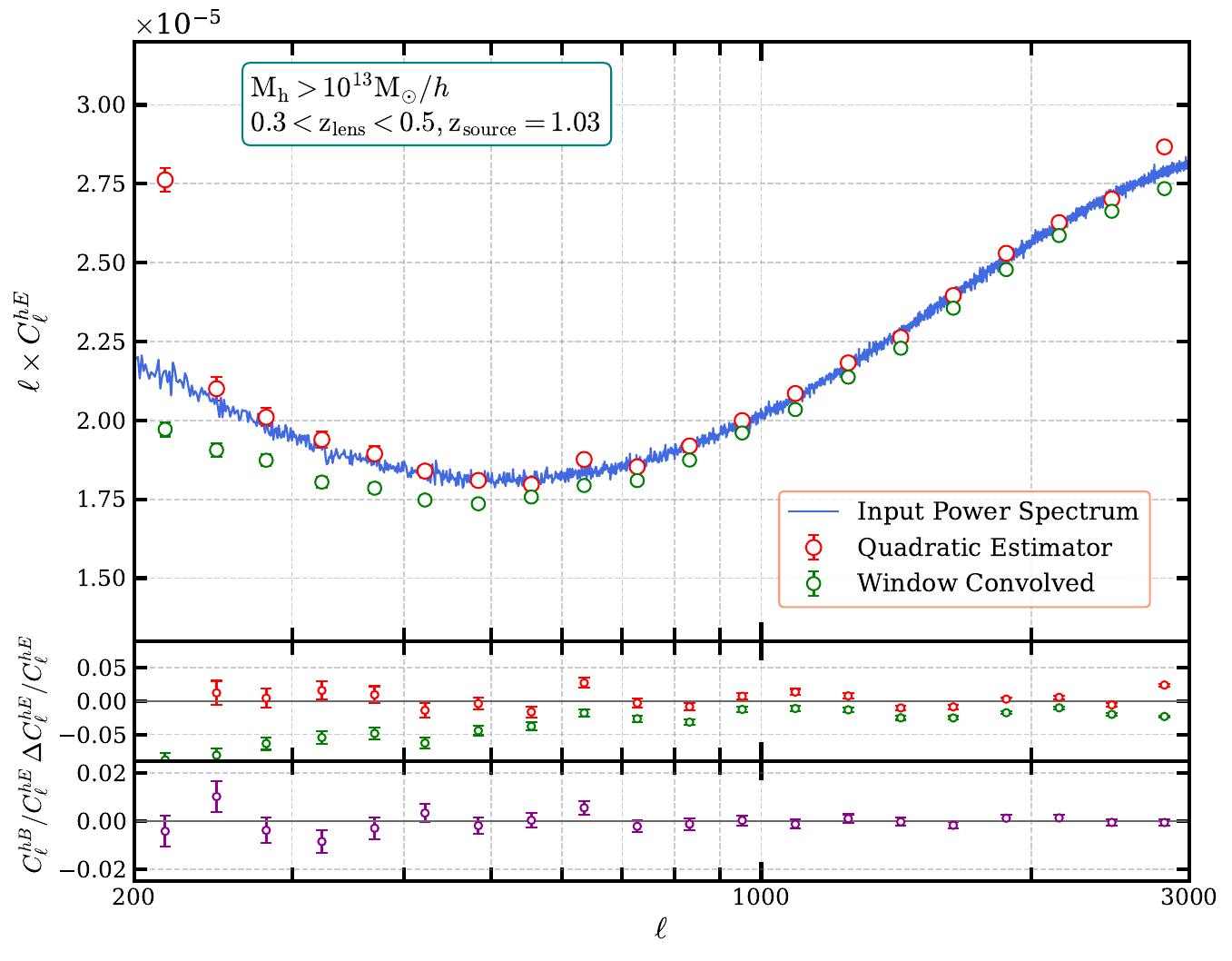}
\caption{{\it Upper panel}: The angular power spectrum $C_{{\rm h}E}(\ell)$  measured by applying the quadratic estimator (Eq.~\ref{eq:gglensing_quadratic_estimator}) to the simulated maps of lens (halos as proxies for lens galaxies) and source galaxies (weak lensing shear) for 100~sq. degrees area as shown in Fig.~\ref{fig:window_fisher}.
The red circle points show the mean of the measured band powers in each multipole bin, obtained from 972 realizations of the simulated maps (Fig.~\ref{fig:window_fisher}), while the error bar in each bin represents the $1\sigma$ uncertainty on the mean. 
Note that the simulated maps include contamination from the simulated intrinsic ellipticities of source galaxies (see text for details).
For comparison, the blue line shows the power spectrum measured from 108 realizations of the original full-sky simulations, while the green circles show the window-convolved power spectrum (see text for the details).
{\it{Middle}}: The fractional difference of $C_{{\rm h}E}$ (red circles) relative to the full-sky $C_{{\rm h}E}$ (blue line in the upper panel, and the denominator in the ratio used here). 
{\it{Lower}}: The non-lensing $B$-mode power spectrum, $C_{{\rm h}B}(\ell)$, measured from the same simulated maps, compared
to $C_{{\rm h}E}(\ell)$.
}
\label{fig:validation}
\end{figure}

In this section, 
we present the main results of this paper: the accuracy of the quadratic estimator in measuring $C_{{\rm g}E}(\ell)$, as estimated using the simulated maps.
Here we used halos as proxies of galaxies in the simulation maps, as described in Section~\ref{subsec:simulation_maps}, and employed the Fisher matrices shown in the {\it right} panel of Fig.~\ref{fig:window_fisher}, which were estimated using 300 realizations of the ancillary $\B{a}$ field. 
For comparison, we also show the {\it{true}}  band powers measured from the original full-sky simulations using the \texttt{healpy} code, which are not affected by the survey window.

The {\it{upper}} panel of Fig.~\ref{fig:validation} shows the band powers of $C_{{\rm h}E}(\ell)$ in each $\ell$-bin, estimated from the simulation maps (Fig.~\ref{fig:window_fisher}) using the quadratic estimator (Eq.~\ref{eq:gglensing_quadratic_estimator}). 
Note that we use the subscript ``h'' to denote halos, which serve as proxies of lens galaxies in our study.
The error bar in each multipole bin denotes the $1\sigma$ statistical error on the mean band power, computed as the standard deviation divided by the square root of the total number of realizations, $\sqrt{972}$. 
The figure shows that the band powers in each $\ell$-bin are consistent with the true angular power spectrum (blue line) to within a few percent accuracy, except for the first and last bins, as shown in the {\it{middle}} panel. 
We note that the band powers in the first and last $\ell$-bins  are not accurate because the Fisher matrices cannot correct for the cross-correlations between these edge bins and multipoles outside the range of bins considered here. 
For further comparison, the green points are the window-convolved band powers, which are corrected by the area fraction of the masked area (with $W=0$) within the FFT area, and significantly deviate from the underlying band powers.

The bottom panel shows the non-lensing $B$-mode band powers, $C_{{\rm h}B}(\ell)$, relative to true $C_{{\rm h}E}(\ell)$.
It is clear that the $B$-mode band powers in each $\ell$-bin are consistent with zero to within the error bar and the leakage of the $E$-mode is below the percent level.

\begin{figure}[t]
\centering 
\includegraphics[width=0.96\textwidth]{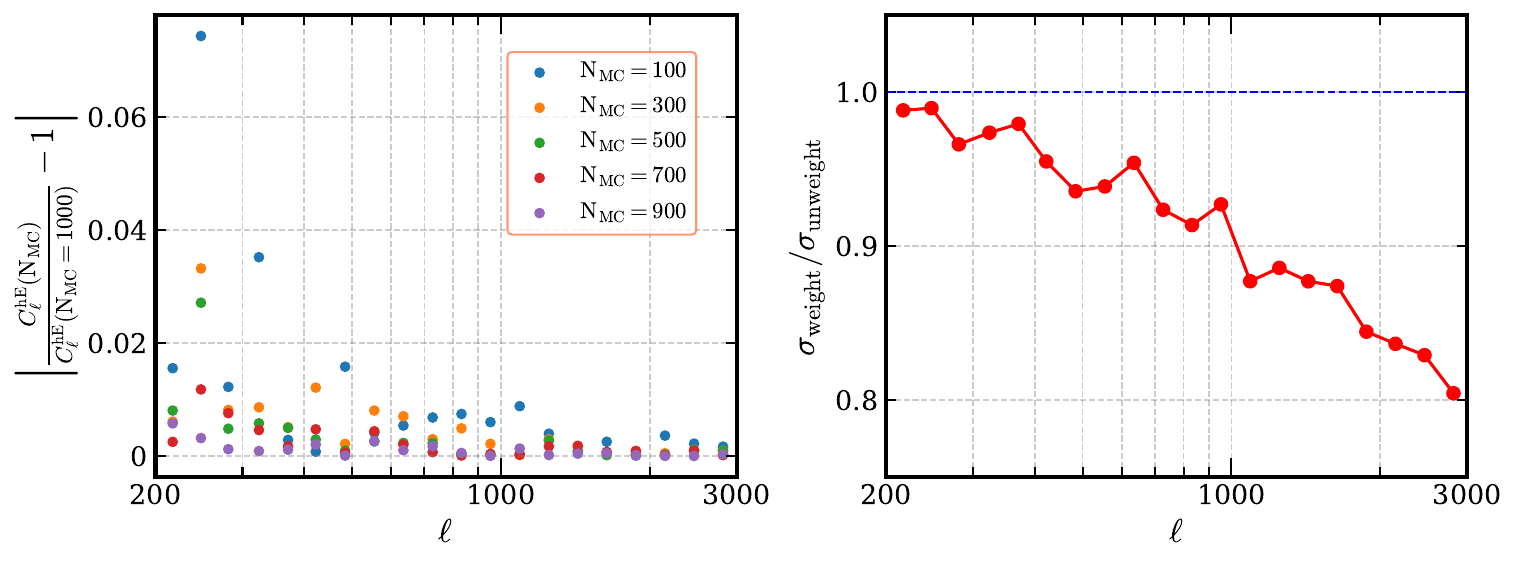}
\caption{{\it Left panel}: The quadratic estimator requires the use of Fisher matrix to correct for the survey window effect when measuring the power spectra, as shown in Fig.~\ref{fig:validation}. 
In this paper we use the Monte Carlo method to estimate the Fisher matrix using a sufficient number of the realizations of random Gaussian ancillary field $\B{a}$.
This plot shows the convergence of the estimated band power in each multipole bin, as a function of the number of the $\{\B{a}\}$-field realizations, denoted as $N_{\rm MC}$. We show the results for $N_{\rm MC}=100, 300, 500, 700$ and $900$, each relative to the result with $N_{\rm MC}=1000$, respectively.
{\it Right}: Comparison of the statistical uncertainties of the band powers measured using the quadratic estimator, with and without the inverse variance weighting of the shape noise (see below Eq.~\ref{eq:shear_estimation}), for each multipole bin. 
} 
\label{fig:fisherconverge}
\end{figure}
The Fisher matrix needs to be evaluated numerically using ancillary field $\{\B{a}\}$~(see around Eq.~\ref{eq:fisher_MonteCarlo_fft}). 
The {\it{left}} panel of Fig.~\ref{fig:fisherconverge} shows the convergence of the power spectrum estimation as a function of the number of realizations of the ancillary field used; to be more precise, we show the results for $\mr{N_{MC}}=100$, 300, 500, 700, 900, or 1000, respectively.
The figure shows that the power spectrum estimation converges to better than the percent level when more than 300 realizations are used, for the simulated shear and halo fields we use in this paper.
Most importantly, for a given data, we can easily (and should) check the convergence of the power spectrum estimation with respect to the number of realizations of the ancillary field used.

In the {\it right} panel of Fig.~\ref{fig:fisherconverge}, we compare the statistical error of the band power measured using the quadratic estimator, with and without the inverse variance weighting of shape noise (see below Eq.~\ref{eq:shear_estimation}), for each multipole bin. 
The reduction in statistical errors due to the weighting is greater at higher multipoles, reaching approximately $20\%$ at $\ell\sim 3000$, which is equivalent  to increasing the survey area by 40\%.

\section{An application of quadratic $C_{{\rm g}E}(\ell)$ estimator: an estimator for the 3D galaxy-matter cross power spectrum, $P_{\rm gm}(k)$}
\label{sec:application}

\subsection{Estimator for $P_{\rm gm}(k)$}

The angular power spectrum, $C_{{\rm g}E}(\ell)$, measured from galaxy-galaxy weak lensing, arises from the line-of-sight projection of the underlying 3D galaxy-matter power spectrum, $P_{\rm gm}(k)$, weighted by the lensing efficiency~(see Eq.~\ref{eq:cl_cge}). 
Therefore, a given multipole ($\ell$) arises from a superposition of different 3D comoving wavenumbers~($k$'s). Here we develop a method to estimate the 3D galaxy-matter power spectrum, $P_{\rm gm}(k)$, from galaxy-galaxy weak lensing measurements.

For a narrow redshift range of lens galaxies, the 3D galaxy-matter power spectrum is expressed in terms of the angular $E$-mode power spectrum in Eq.~(\ref{eq:cl_cge}) as 
\begin{align}
P_{\rm gm}(k;\bar{\chi}_L)\simeq \frac{\Sigma_{\mr{crit}}(\bar{\chi}_L)}{\rho_{\mr{crit}}\Omega_m}
\bar{\chi}_L^2C_{{\rm g}E}(\ell=\bar{\chi}_L k),
\label{eq:Pgm_est}
\end{align}
where $\bar{\chi}_L$ is the average of comoving angular diameter distances to lens galaxies within the narrow redshift range, $C_{{\rm g}E}$ is the window-free $E$-mode power spectrum estimated using the quadratic estimator developed in this paper.
To be precise, galaxy-galaxy weak lensing probes the power spectrum of wavenumbers perpendicular to the line-of-sight direction, and we can assume that $C_{{\rm g}E}(\ell)$ gives an estimate of the 3D power spectrum $P_{\rm gm}(k)$ under 
the assumption of statistical isotropy. 

Let us consider a situation where we have a sample of lens galaxies over some range of lens redshifts ($z_L$) and a sample of source galaxies at $z_s$ satisfying $z_s>z_L$.
The procedures we consider for measuring $P_{\rm gm}(k)$ are as follows. i) We divide the lens samples into multiple redshift slices. The width of each lens slice is chosen to be sufficiently large so that correlations between the different lens slices can be neglected; e.g., a width thicker than $\sim 30h^{-1}{\rm Mpc}$ would satisfy this condition. 
ii) We measure the angular power spectrum, $C_{{\rm g}E}(\ell;z_L)$ for each lens slice using the quadratic estimator developed in this paper. We then convert the measured $C_{{\rm g}E}$ to $P_{\rm gm}$ using Eq.~(\ref{eq:Pgm_est}) in each slice.
iii) We combine the estimated $P_{\rm gm}$ in each lens slice to estimate the averaged $P_{\rm gm}$ over all lens slices. Therefore, the estimator for the average $P_{\rm gm}$ can be given as
\begin{align}
\hat{P}_{\rm gm}(k_b)=\frac{1}{\rho_{\rm crit}\Omega_{\rm m}\sum_L w_L}\sum_{L} w_L \Sigma_{\rm crit}(\bar{\chi}_L)\bar{\chi}_L^2
\left.\hat{C}_{{\rm g}E}(\ell)\right|_{\ell \in \bar{\chi}_Lk_b},
\label{eq:Pgm_estimator}
\end{align}
where the summation runs over multiple lens slices, $w_L$ is the weight for the $L$-th lens slice, $k_b$ is the $k$-bin we want to measure, and $\ell \in \bar{\chi}_L k_b$ indicates that we combine all multipole data that fall within the $k$-bin through the conversion between angular and 3D scales ($\ell\leftrightarrow k)$ in each slice. 

Using a similar method to that in Refs.~\cite{2000ApJ...540..605P,2005MNRAS.361.1287M,2018MNRAS.478.4277S} and assuming the shot noise limited regime for lens sample, we can find the inverse-variance weighting, given as 
\begin{align}
w_L=\frac{f_{A_{L}}\bar{n}^{\rm 2D}(z_L)}{\Sigma_{\rm crit}(\bar{\chi}_L)^2},
\label{eq:Pgm_optimal_weight}
\end{align}
where $\bar{n}^{\rm 2D}(z_L)$ is the average surface number density of lens galaxies in the $L$-th lens slice (with units of $[(h^{-1}{\rm Mpc})^{-2}]$), and $f_{A_L}$ is the area fraction of the data in the $L$-th lens slice relative to the Fourier area (see Fig.~\ref{fig:pgm_window_slices}).
The above weight is derived from the fact that the variance of the projected number density fluctuation field of galaxies is given by $\sigma^2(\Delta_{\rm g})\propto 1/[S_L\bar{n}^{\rm 2D}(z_L)]$, where $S_L$ is the area of the data for the $L$-th lens slice. Note that $S_L$ is the area where data exists, excluding the zero-padded and masked regions.
We will show that this weighting can indeed improve the statistical precision of $P_{\rm gm}(k)$ measurement using the simulation data. Note that the weighting for galaxy-galaxy weak lensing (galaxy shape) measurements is accounted when measuring $C_{{\rm g}E}(\ell)$ (see below Eq.~\ref{eq:shear_estimation} and the {\it right} panel of Fig.~\ref{fig:fisherconverge}). 

\subsection{Validation of $P_{\rm gm}$ estimator}
\label{sec:validation_Pgm_estimator}

\begin{figure}[htbp]
\centering 
\includegraphics[width=1\textwidth]{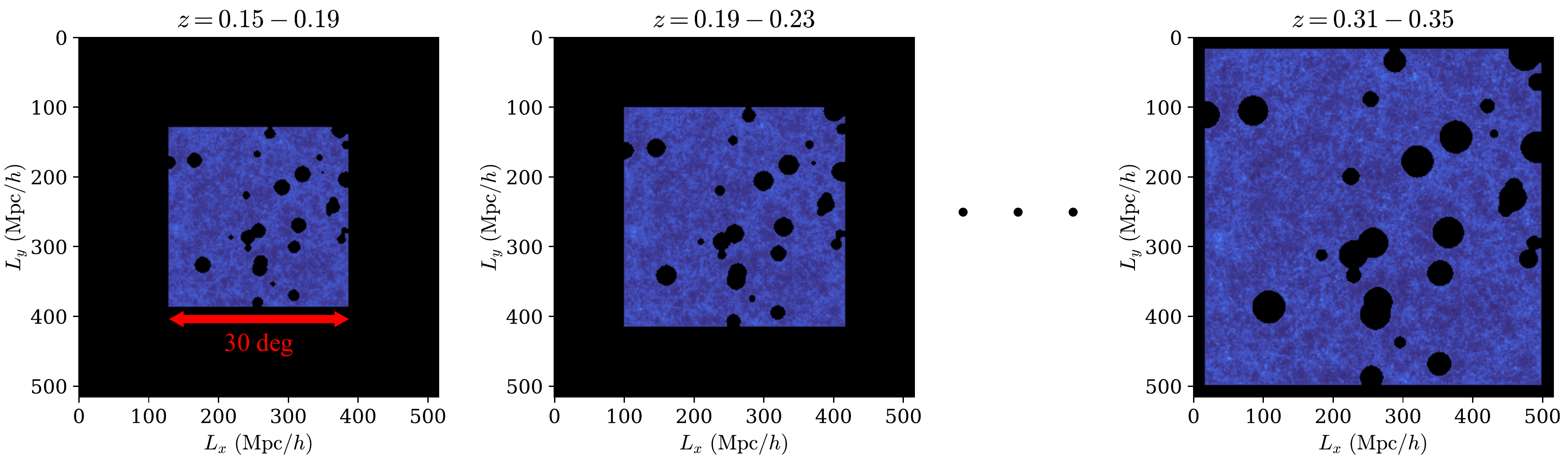}
\caption{An example of a setup for estimating the 3D galaxy-matter cross-power spectrum from galaxy-galaxy weak lensing measurements. Here we consider a survey region of $30\times 30$~sq. degrees, a lens sample consisting of halos with masses greater than $10^{13}h^{-1}M_\odot$ in the redshift range $0.15<z<0.35$, which are taken as proxies for the BOSS LOWZ galaxies, and a sample of source galaxies at $z_s=1.03$.
We divide the lens halos into 5 slices, each of which has the same projected area of about $516^2~(h^{-1}{\rm Mpc})^2$ including the zero-padded region and has a radial thickness of $107~h^{-1}{\rm Mpc}$. 
As can be found, we include masked regions, e.g. due to bright stars. The color indicates the lensing convergence field ($\kappa$) in each lens plane (therefore all the color maps are the same, although they are scaled by different factors). Note that, for the lens sample, we employ different survey masks as described in Section~\ref{subsec:simulation_maps}.
}
\label{fig:pgm_window_slices}
\end{figure}
\begin{figure}[htbp]
\centering 
\includegraphics[width=1\textwidth]{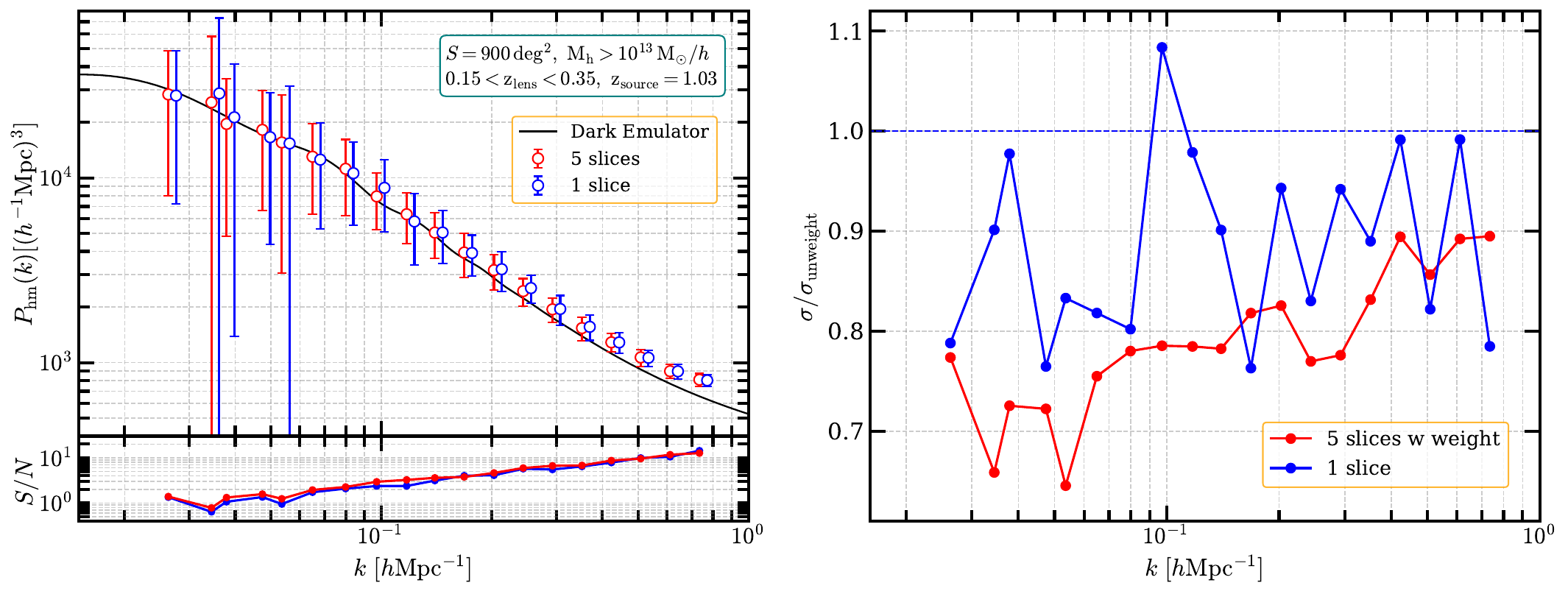}
\caption{{\it Upper-left panel}: The red points with error bars show the 3D halo-matter cross-power spectrum, $P_{\rm hm}(k)$, estimated using the quadratic estimator (Eq.~\ref{eq:gglensing_quadratic_estimator}) for the galaxy-galaxy weak lensing power spectrum in each redshift slice of the lens sample, based on the setup shown in Fig.~\ref{fig:pgm_window_slices}. 
The error bar in each $k$ bin denotes the $1\sigma$ statistical uncertainty for a survey area covering $30\times 30$~sq. degrees. For comparison, the blue points denote similar results using a single slice of the lens sample. Although we use the same wavenumber bins for both the 5 slices and single slice cases, we slightly shift the blue points to the right for easier comparison.
The solid curve shows the \texttt{Dark~Emulator} prediction (see text for details).
{\it Lower-left}: The signal-to-noise ratio for the band power estimation in each $k$ bin.
{\it Right}: The red points show the improvement in the error bars in each $k$-bin when using the weight defined in Eq.~(\ref{eq:Pgm_optimal_weight}) for the 5 lens slices, compared to the case without weighting in the 5-slice setup. For comparison, the blue points show the results for the single lens slice case, also compared to the unweighted 5-slice case.} 
\label{fig:measuredPhm}
\end{figure}

In this section, we show validation of the $P_{\rm gm}(k)$ estimator using the simulated maps of the halo field, as proxies of galaxies, and weak lensing shear. 
We use the same full-sky simulations described in Section~\ref{subsec:simulation_maps} to construct the simulated maps of halos and weak lensing shear. Here we extract $30^\circ\times 30^\circ(=900~{\rm deg.}^2)$ rectangular-shaped region from each of the 108 realizations. We use halos with masses greater $10^{13}h^{-1}M_\odot$ in the redshift range $0.15<z<0.35$, and use the shear field at $z_s=1.03$. The area of 900~sq. degrees corresponds to an area of the upcoming full Subaru HSC data ($\sim 1,100$~sq. degrees) \citep{aihara_hsc_overview}, and the halo mass and redshift range of lens halos roughly correspond to the typical halo mass of the LOWZ sample \citep{2013AJ....145...10D,reid_sdss-iii_2015} \citep[also see Ref.][for the typical halo mass of LOWZ galaxies estimated using the weak lensing measurements]{2022PhRvD.106h3520M}. The source redshift is a typical mean redshift for source galaxies such as that for the HSC data. Note that we also include intrinsic ellipticities to weak lensing shear, as described in Section~\ref{subsec:simulation_maps}.

In Fig.~\ref{fig:pgm_window_slices} we show how to divide lens samples in the redshift range ($0.15<z<0.35$) into 5 lens slices, as an example. Each lens slice has the same radial thickness of the comoving length, about $107~h^{-1}{\rm Mpc}$, and we construct the projected number density field of halos (as proxies of galaxies) in each lens slice. 
For each lens slice, we need to convert the angular scales to the comvoing projected length scales using $\Delta\B{\chi}_\perp =\bar{\chi}_L\Delta\B{\theta}$, where $\bar{\chi}_L$ is the comoving angular diameter distance for the mean redshift of the $L$-th lens slice. 

To do this, we need to assume the reference cosmology, e.g. $\Lambda$CDM model, when converting the angular scales to the comoving scales, where the reference cosmology generally differs from the underlying true cosmology. 
As illustrated in Fig.~\ref{fig:pgm_window_slices}, each of the lens slices has the same area, $516^2~(h^{-1}{\rm Mpc})^2$ in this example, including the zero-padded region. 
By performing this conversion, we can measure the same (projected) wavenumber of $k_\perp$ across the multiple lens slices.  
For this reason, the area including the data and zero-padded regions is determined by the maximum lens-redshift slice. 
We also note that the use of the reference cosmology, which differs from the true cosmology, causes systematic effects in the measured power spectrum. However, this geometrical effect can be easily corrected for, e.g., using the method in Refs.~\cite{more_2013ApJ...777L..26M,2022PhRvD.106h3520M}. In this study, we simply use the same cosmological models used in the ray-tracing simulation.

The {\it{left}} panel of Fig.~\ref{fig:measuredPhm} shows the result for the $P_{{\rm gm}}(k)$ measurement. 
As described above, we first measure the angular power spectrum $C_{{\rm g}E}(\ell; z_L)$ in each of the 5 lens slices using the quadratic estimator for galaxy-galaxy weak lensing. 
We then combine the measured $C_{{\rm g}E}$'s to estimate the galaxy-matter cross power spectrum, more exactly here the halo-matter cross spectrum $\hat{P}_{\rm hm}(k)$. 
Here, we use 19 logarithmic bins in the wavenumber range $k~=~[0.024,0.8]~h{\rm Mpc}^{-1}$. 
Since we use the same 2D Fourier area as given in Fig.~\ref{fig:pgm_window_slices}, we can combine the same Fourier modes, which fall within a given $k$-bin, to measure $P_{\rm hm}(k)$.
The figure shows the measured $P_{\rm hm}(k)$ in each $k$-bin, where the error bars represent the statistical uncertainty on the mean band powers estimated from the 108 realizations. 
Note that the measured $P_{\rm hm}(k)$ is free from the redshift space distortion effect, as galaxy-galaxy weak lensing can measure only Fourier modes perpendicular to the line-of-sight direction. The figure also shows the result when only the single slice for $0.15<z<0.35$ is used. 
Here, we set the same comoving area with 5 slices case to obtain the same wavenumber. (Note that we slightly shift the single slice plots in this figure for illustrative purposes.)
Although the band power in each $k$ bin fairly well agree with that for the 5 slices result, the error bar is larger. 

The ray-tracing simulations of Ref.~\cite{takahashi_full-sky_2017} do not provide the dark matter distribution in each lens redshift slice, although the halo catalog is available. Therefore, we compare the estimated $P_{\rm hm}(k)$ with the prediction computed using the \texttt{Dark~Emulator} code that is given by Ref.~\cite{nishimichi_dark_2019}\footnote{The \texttt{Dark~Emulator} and the detailed information are found in \url{https://darkquestcosmology.github.io}.}. 
The \texttt{Dark~Emulator} provides predictions of the real-space halo-matter power spectrum for a given input of cosmological parameters within the flat $\Lambda$CDM model, as well as halo mass threshold and redshift. 
Here we assumed the cosmological models used in the ray-tracing simulation, and adopted $M>10^{13}h^{-1}M_\odot$ and $z=0.28$ for the weighted mean of lens redshifts.
The measured $P_{\rm hm}(k)$ fairly well agrees with the model prediction. 
The lower panel shows the signal-to-noise ratio for the band power measurement in each $k$ bin. It is clear that an HSC-like survey will allow us to measure the band power with high significance at $k\gtrsim 0.1~h{\rm Mpc}^{-1}$.

The {\it right} panel of Fig.~\ref{fig:measuredPhm} compares the error bars of the $P_{\rm hm}$ band power estimation in each $k$ bin, with and without the use of the weight $w_L$ (Eq.~\ref{eq:Pgm_optimal_weight}). 
The figure shows that using the weight leads to an improvement in the error bars in each $k$-bin by up to 20-30\%. The figure also shows that dividing the lens halos and using a proper weight yields improved error bars in the estimation of $P_{\rm hm}$ compared to the single slice case. 
We verified that using 10 lens redshift slices yields results that are nearly identical to those obtained with 5 slices.

\section{Conclusion and Discussion}
\label{sec:Conclusion}

In this paper, we developed the quadratic estimator to measure the $E$- and $B$-mode power spectra of galaxy-galaxy weak lensing, $C_{{\rm g}E}(\ell)$ and $C_{{\rm g}B}(\ell)$. 
Our estimator can properly correct for the survey window effects as well as minimize the leakage of $E$-mode into the $B$-mode.
To enable efficient computation, our method employs FFTs on pixelized maps of the lens galaxy distribution and source galaxy ellipticities under the flat sky approximation, and uses a sufficient number realization of ancillary Gaussian fields to estimate the Fisher matrix via the Monte Carlo method. Then the inverse of the Fisher matrix is used to deconvolve the survey window effects. 
We also showed in Appendix~\ref{subsec:correspondence} that the quadratic estimator becomes equivalent to the pseudo-$C_\ell$ estimator when a simple form of the weight is used. 

Using the shear fields and halo catalogs in the ray-tracing simulations, we showed that the quadratic estimator can recover the underlying $E$-mode power spectrum to within a few percent in fractional error, while minimizing the leakage of $E$-modes into the $B$-mode power spectrum (Fig.~\ref{fig:validation}). 
As an application of this estimator, we also developed a method to measure the 3D galaxy-matter power spectrum, $P_{\rm gm}(k)$, by dividing lens galaxies into multiple redshift slices (Fig.~\ref{fig:measuredPhm}). 
Applying the optimal weight on each slices in the shot/shape noise limited regime, we found that the error bars can be reduced by about 20--$30\%$ compared to the case without weighting. 

The method developed in this paper has several applications. 
First, the quadratic estimator can be straightforwardly applied to measure the $E$- and $B$-mode auto-power spectra of cosmic shear. 
Since the quadratic estimator is derived based on the maximum likelihood of the data vector, it would be interesting to develop an optimal quadratic estimator to measure the cosmic shear power spectra using the inverse of the covariance matrix, $C^{-1}$, instead of the diagonal $H$ matrix in Eq.~(\ref{eq:estimator_general}). 
If we can improve the statistical precision of the power spectrum estimation over the entire range of multipoles, both in the sample variance and shot noise dominated regimes, it is equivalent to having a survey with wider area coverage and higher number density.
Secondly, the maximum likelihood approach can be applied to develop a window-free estimator for weak lensing bispectra, such as  shear-shear-shear, galaxy-galaxy-shear and shear-galaxy-galaxy bispectra.  
This can be done by extending the methods in Refs.~\citep{philcox_cosmology_2021-1,philcox_optimal_2023,philcox_optimal_2023-1} for galaxy survey data. Making use of the spin-2 nature of weak lensing shear, 
the weak lensing bispectra, if measured, offer a novel approach to searching for parity-violating signals in  large-scale structures.  
In addition, the $B$-mode bispectra can be used as additional diagnostics of systematic effects, similar to the $B$-mode power spectrum.

The 3D galaxy-matter power spectrum, $P_{\rm gm}(k)$, measured from galaxy-galaxy weak lensing, can be used to observationally constrain the galaxy bias of the lens galaxy sample. 
Therefore, it would be interesting to explore cosmology inference by jointly combining $P_{\rm gm}(k)$ with the auto-power spectrum of the lens galaxies, $P_{\rm gg}(k)$, since most of the previous works are based on the real-space statistics \citep{2020PhRvD.102h3520S,2022PhRvD.106h3520M,miyatake_hyper_2023,2023PhRvD.108l3521S}.
In particular, it is worth exploring cosmology inference by combining $P_{\rm gm}(k)$ with the redshift-space power spectrum -- namely, by extending the full-shape cosmology analysis of the redshift-space power spectrum to jointly include $P_{\rm gm}(k)$, for a spectroscopic sample of galaxies. 
In such analyses, one can use the EFTofLSS model \citep{philcox_boss_2022} or an emulator-based model \citep{2022PhRvD.105h3517K} to self-consistently model both $P_{\rm gg}$ and $P_{\rm gm}$ in cosmological inference.

\appendix

\section{Equivalence to the pseudo-$C_\ell$ method} \label{subsec:correspondence}
In this section, we examine the equivalence between the quadratic estimator and the pseudo-$C_\ell$ method, the latter being a widely used method in the literature~\citep{hikage_shear_2011,hikage_pseudo-spectrum_2016}.

The pseudo-$C_\ell$ method estimates the underlying power spectrum by deconvolving the window function effect from the measured power spectrum as follows. The measured power spectrum, $P^W(k)$, is involved with the survey window function:
\begin{align}
    P^{W}(k) &=\int_{\B{k'}}|W(\B{k}-\B{k}')|^2P^{\rm{true}}(k')~,
    \label{eq:pseudo_cl}
\end{align} 
where $W(\B{x})$ is the window function.
Then, the power spectrum is discretized into a finite number of $k$-bins, which are intended to be measured:
\begin{align}
    P^W_\alpha &= \sum_{\beta} {\cal M}_{\alpha\beta}P^{\rm{true}}_{\beta},
    \label{eq:pseudo_cl_convolved}
\end{align}
where the indices $\alpha, \beta$ denote the $\alpha$- or $\beta$-th $k$ bin, and ${\cal M}_{\alpha\beta}$ is the matrix describing the discretized window effect, defined as
\begin{align}
{\cal M}_{\alpha\beta}&\equiv \int_{\B{k}}\Theta_\alpha(\B{k})\int_{\B{k'}} \Theta_\beta(\B{k}')|W(\B{k}-\B{k'})|^2.
\label{eq:pseudo_fisher}
\end{align}
Here $\Theta_\alpha(\B{k})$ is the step function used in Eq.~(\ref{eq:cov_derivative}).
Then the band power in the $\alpha$-th bin can be estimated as 
\begin{align}
\hat{P}_\alpha &= {\cal M}^{-1}_{\alpha\beta}P^W_\beta.
\label{eq:pseudo_cl_estimator}
\end{align}

Let us return to the quadratic estimator. In the following, we consider the simplified estimator for Eq.~(\ref{eq:estimator_general}) to demonstrate the equivalence with the pseudo-$C_\ell$ method. 
To do this, we adopt the diagonal matrix for $H$, i.e., $H(\B{x}_1,\B{x}_2)=\delta^K_{\B{x}_1,\B{x}_2}$, where $\delta^K_{\B{x}_1,\B{x}_2}$ is the Kronecker delta function.
In this case, the Fisher matrix is computed as
\begin{align}
F_{\alpha\beta}
&= \frac{1}{2}\sum_{\B{x}_1,\B{x}_2} C_{,\alpha}(\B{x}_1,\B{x}_2)C_{,\beta}(\B{x}_2,\B{x}_1)
\nonumber\\
&= \frac{1}{2}\sum_{\B{x}_1,\B{x}_2} W(\B{x}_1)W(\B{x}_2)
\int_{\B{k}}\Theta_{,\alpha}(\B{k})e^{i\B{k}\cdot(\B{x}_1-\B{x}_2)}
\int_{\B{k}'}\Theta_{,\beta}(\B{k}')e^{i\B{k}'\cdot(\B{x}_2-\B{x}_1)}\nonumber\\
&\propto \frac{1}{2}\int_{\B{k}}\Theta_{,\alpha}(\B{k})\int_{\B{k}'}\Theta_{,\beta}(\B{k}')
\int_{\B{x}_1} W(\B{x}_1) e^{i(\B{k}-\B{k}')\cdot\B{x}_1}
\int_{\B{x}_2} W(\B{x}_2) e^{-i(\B{k}-\B{k}')\cdot\B{x}_2}\nonumber\\
&= \frac{1}{2}\int_{\B{k}}\Theta_{,\alpha}(\B{k})\int_{\B{k}'}\Theta_{,\beta}(\B{k}')
\left|W(\B{k}-\B{k}')\right|^2~ .
\label{eq:simplefisher}
\end{align}
In the third line on the right-hand side of the above equation, we approximate the discrete sum by an integral over the FFT region that contains the survey area.
In the same manner, one can prove that the data term in Eq.~(\ref{eq:estimator_general}) is actually the same as the window convolved band-power $P^W_\alpha$.
Therefore, comparing Eqs.~(\ref{eq:pseudo_fisher}) and (\ref{eq:pseudo_cl_estimator}) with Eq.~(\ref{eq:estimator_general}), we find that the quadratic estimator becomes equivalent to the pseudo-$C_\ell$ estimator, {\it if} we assume that $H$ is a diagonal matrix and the window function satisfies $W(\B{x})^2=W(\B{x})$. 
This condition is indeed satisfied when $W(\B{x})=1$ or 0.

\acknowledgments
We would like to thank Oliver Philcox for very useful discussion in the early phase of this project and also 
thank Ryo~Terasawa, Shintaro~Nakano
and Ryuichi~Takahashi for useful discussion. This work was supported in part by
JSPS KAKENHI Grant Number 23KJ0747, and by World Premier International Research Center Initiative (WPI Initiative), MEXT, Japan.

\bibliographystyle{JHEP}
\bibliography{refs}
\end{document}